\documentclass[a4paper,11pt]{article}
\usepackage{pos}

\title{Effect of eigenstates on spectra in coupled-channel scattering 
with the chiral unitary model}
\ShortTitle{Effect of eigenstates on spectra in coupled-channel scattering}

\author*[a]{Takuma Nishibuchi}
\author[a]{Tetsuo Hyodo}

\affiliation[a]{Tokyo Metropolitan University,\\
  1-1, Minamiosawa,Hachioji, Tokyo, Japan}

\emailAdd{nishibuchi-takuma@ed.tmu.ac.jp}
\emailAdd{hyodo@tmu.ac.jp}

\abstract{
In recent years, as experimental data on the excited $\Xi(1620)$ and $\Xi(1690)$ states have accumulated, theoretical analyses based on chiral dynamics have also been actively pursued. In this study, we construct theoretical models within the chiral unitary approach to reproduce recent experimental data, and perform a model interpolation to examine the relationship between the poles appearing in different models. We then calculate the invariant mass spectra of the $\pi^+\Xi^-$ system in the $\Xi_c \to \pi\pi\Xi$ decay, using the scattering amplitudes obtained from each model, to investigate how the distinct pole structures influence the observable spectrum.}

\FullConference{The 21st International Conference on Hadron Spectroscopy and Structure (HADRON2025)\\
 27 - 31 March, 2025\\
Osaka University, Japan\\}

\begin{document}
\maketitle

\section{Introduction}

In recent years, the excited $\Xi$ baryons, particularly $\Xi(1620)$ and $\Xi(1690)$, have attracted increasing attention due to the availability of new experimental data~\cite{ParticleDataGroup:2024cfk,Guo:2017jvc,Hyodo:2020czb}. For instance, in 2019, the Belle collaboration observed peaks corresponding to $\Xi(1620)$ and $\Xi(1690)$ in the $\pi^+\Xi^-$ invariant mass distribution of the $\Xi_c \rightarrow \pi\pi\Xi$ decay~\cite{Belle:2018lws}. In addition, the ALICE collaboration determined the $K^-\Lambda$ scattering length from correlation functions obtained via femtoscopy in Pb-Pb collisions in 2021~\cite{ALICE:2020wvi}. These developments have been accompanied by active theoretical investigations~\cite{Ramos:2002xh,Garcia-Recio:2003ejq,Sekihara:2015qqa,Khemchandani:2016ftn,Feijoo:2023wua,Nishibuchi:2023acl,Li:2023olv,Sarti:2023wlg,Feijoo:2024qqg,Nishibuchi:2025uvt}.

In our previous work~\cite{Nishibuchi:2023acl}, we constructed two theoretical models for the $\Xi(1620)$: Model 1, based on the mass and decay width reported by the Belle experiment, and Model 2, based on the $K^-\Lambda$ scattering length determined by the ALICE experiment. However, the poles obtained from these two models suggested completely different eigenstates. The pole in Model 1 corresponds to a quasibound state located slightly below the $\bar{K}\Lambda$ threshold, whereas the pole in Model 2 lies above the threshold and resides on a different Riemann sheet from that of Model 1. It should be noted that Ref.~\cite{Nishibuchi:2023acl} focused on the calculation of meson--baryon two-body scattering amplitudes, while the evaluation of three-body decay processes, such as the experimentally observed $\Xi_c \rightarrow \pi\pi\Xi$ decay, was not conducted.

Although poles corresponding to the $\Xi(1620)$ have been found in both Model 1 and Model 2, it remains unclear whether they originate from the same physical state or from distinct sources. Moreover, to perform an analysis that is more directly comparable to experimental observations, it is essential to investigate how these pole structures affect the three-body decay spectrum. In this study, we examine the influence of different pole structures on the $\pi\pi\Xi$ invariant mass distribution in the $\Xi_c \rightarrow \pi\pi\Xi$ decay. For further details, we refer the reader to Ref.~\cite{Nishibuchi:2025uvt}.

\section{Formulation}

In this study, we employ the chiral unitary approach~\cite{Kaiser:1995eg,Oset:1997it,Oller:2000fj,Hyodo:2011ur,Hyodo:2020czb,Nishibuchi:2023acl}, which dynamically generates the $\Xi(1620)$ resonance through chiral meson-baryon interactions. Specifically, the meson-baryon scattering amplitude $T_{ij}(W)$ at total energy $W$ is obtained by solving the coupled-channel scattering equation:
\begin{align} \label{eq:tscat}
T_{ij}(W) = V_{ij}(W) + \sum_k V_{ik}(W)\, G_k(W)\, T_{kj}(W),
\end{align}
where $V_{ij}(W)$ denotes the interaction kernel, and $G_k(W)$ is the loop function. For the interaction kernel, we use the Weinberg-Tomozawa term, given by
\begin{align} 
V_{ij}^{\mathrm{WT}}(W) = -\frac{C_{ij}}{4f_i f_j}(2W - M_i - M_j)
\sqrt{\frac{M_i + E_i}{2M_i}} \sqrt{\frac{M_j + E_j}{2M_j}},
\end{align}
where $C_{ij}$ is the group-theoretical coefficient representing the interaction strength between channels, $f_i$ is the meson decay constant, and $M_i$ is the baryon mass. The loop function $G_k(W)$ arises from momentum integrals over intermediate states and contains ultraviolet divergences. These divergences are regularized using dimensional regularization, and $G_k(W)$ becomes a function of the energy $W$ and the subtraction constant $a_k$, given by
\begin{align} \label{eq:gloop}
   &\quad G_k[W;a_k(\mu_{\mathrm{reg}})] \nonumber \\
   &=\frac{2M_k}{16\pi^2}\biggl[a_k(\mu_{\mathrm{reg}})+\ln\frac{m_kM_k}{\mu^2_{\mathrm{reg}}}
   +\frac{M_k^2-m_k^2}{2W^2}\ln\frac{M_k^2}{m_k^2} 
   +\frac{\lambda^{1/2}(W^{2},M_{k}^{2},m_{k}^{2})}{2W^2}\nonumber \\
   &\quad \times\Bigl\{\ln(W^2-m_k^2+M_k^2+\lambda^{1/2}(W^{2},M_{k}^{2},m_{k}^{2}))
   +\ln(W^2+m_k^2-M_k^2+\lambda^{1/2}(W^{2},M_{k}^{2},m_{k}^{2})) \nonumber \\
   &\quad -\ln(-W^2+m_k^2-M_k^2+\lambda^{1/2}(W^{2},M_{k}^{2},m_{k}^{2})) 
   -\ln(-W^2-m_k^2+M_k^2+\lambda^{1/2}(W^{2},M_{k}^{2},m_{k}^{2}))\Bigr\}\biggr],
\end{align}
where $a_k$ effectively serves as a cutoff for channel $k$, $\mu_{\mathrm{reg}}$ is the renormalization scale, $m_k$ is the meson mass, and $\lambda$ denotes the K\"all\'en function.

In our previous study~\cite{Nishibuchi:2023acl}, we proposed two models: Model~1, constructed by adjusting the subtraction constants $a_i$ to reproduce the pole position corresponding to the mass and decay width of the $\Xi(1620)$ resonance reported by the Belle experiment; and Model~2, which reproduces the $K^-\Lambda$ scattering length determined by the ALICE experiment. In the present work, to clarify the relationship between these two models, we perform a model interpolation using the following formula:
\begin{align}
a_i(x) = x\, a''_i + (1 - x)\, a'_i,
\end{align}
where $a'_i$ and $a''_i$ denote the subtraction constants corresponding to Model~1 and Model~2, respectively. By continuously varying the parameter $x$ from 0 to 1, one can smoothly track the transition from Model~1 to Model~2.

Furthermore, in this study, we calculate the decay amplitude of the $\Xi_c \rightarrow \pi\pi\Xi$ process with the aim of comparing our results with experimental data, following Ref.~\cite{Miyahara:2016yyh}. The invariant mass distribution of the $\pi^+\Xi^-$ channel as a function of the invariant mass $M_{\rm{inv}}$ is given by
\begin{align}
\frac{d\Gamma_j}{dM_{\rm{inv}}}
= \frac{1}{(2\pi)^3} \frac{p_{\pi^+} \tilde{p}_j M_j}{M_{\Xi_c^+}}
\left| V_P \left( h_j + \sum_i h_i G_i(M_{\rm{inv}}) T_{ij}(M_{\rm{inv}}) \right) \right|^2,
\end{align}
where $V_P$ corresponds to the weak decay process prior to the final-state interaction and is treated as a constant in this work. The coefficient $h_i$ represents the weight of the intermediate channel $i$, and $T_{ij}$ and $G_i$ are given by Eqs.~\eqref{eq:tscat} and~\eqref{eq:gloop}, respectively. Here, $p_{\pi^+}$ denotes the three-momentum of the $\pi^+$ meson not involved in the final-state interaction, measured in the rest frame of the $\Xi_c^+$, while $\tilde{p}_j$ represents the three-momentum in the center-of-mass frame of the meson-baryon system involved in the final-state interaction.

\section{Results}

\begin{figure}[tbp]
\centering
\includegraphics[width=8cm]{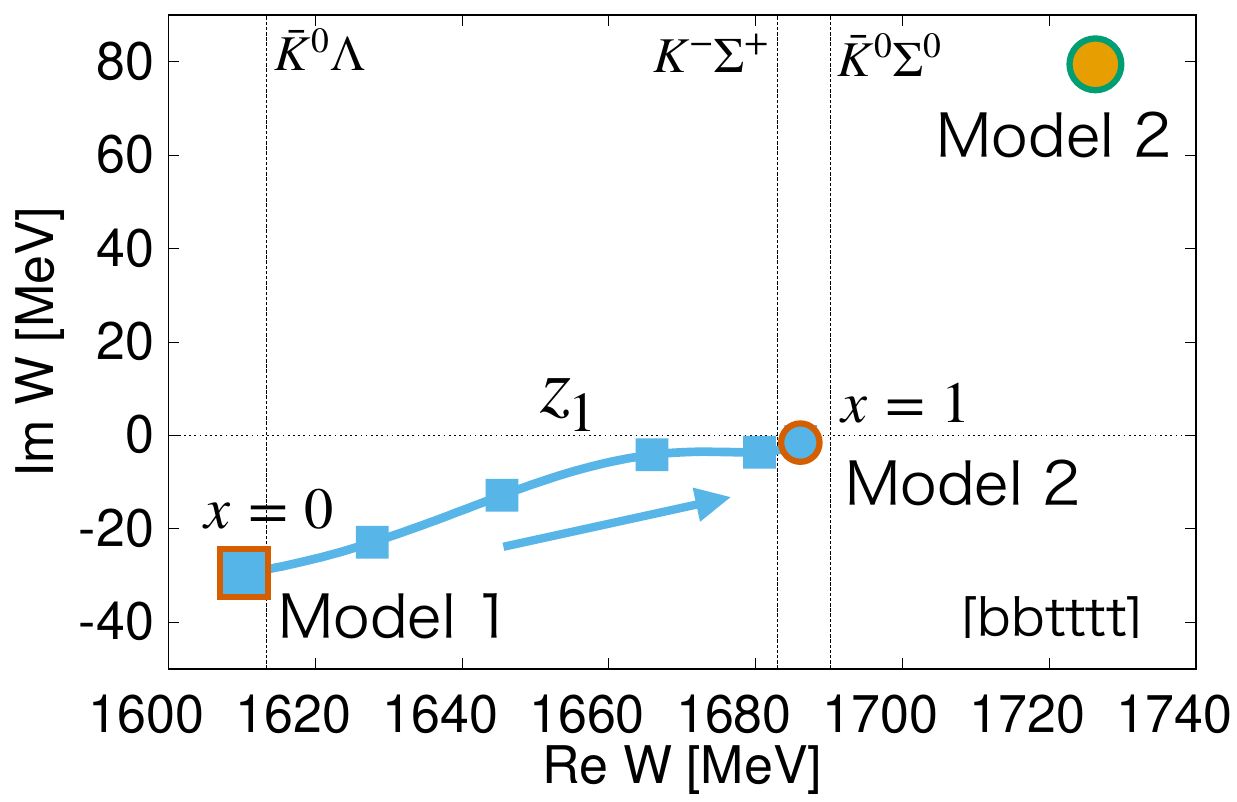}
\caption{Trajectory of the pole $z_1$ obtained through model interpolation. The pole corresponding to Model~1 ($x=0$) is indicated by a square, while those for Model~2 ($x=1$) are shown by circles. Intermediate poles for $x$ varied in steps of 0.2 are represented by squares.}
\label{fig:pole_trajectory}
\end{figure}

In Ref.~\cite{Nishibuchi:2023acl}, it was shown that the pole corresponding to the $\Xi(1620)$ in Model~1, denoted as $z_1$, is located on the [bbtttt] Riemann sheet at $1610 - 30i~\text{MeV}$, while the pole in Model~2, denoted as $z_2$, resides on the [ttbttt] Riemann sheet at $1726 + 80i~\text{MeV}$. Figure~\ref{fig:pole_trajectory} illustrates the positions of these two poles, along with the trajectory of the pole originating from $z_1$ as the subtraction constants $a_i$ are continuously varied via the interpolation parameter $x$, from the values in Model~1 to those in Model~2.

As the values of $a_i$ approach those of Model~2, the real part of the energy of the pole $z_1$ increases. However, the resulting pole position at $x=1$ (i.e., when $a_i$ are set to the values of Model~2) does not coincide with $z_2$. Throughout the deformation process, $z_1$ remains on the [bbtttt] Riemann sheet and never reaches the [ttbttt] sheet where $z_2$ is located. Conversely, even when starting from $z_2$ and performing a similar extrapolation toward the parameter set of Model~1, one cannot reach the position or the Riemann sheet of $z_1$~\cite{Nishibuchi:2025uvt}. These results indicate that the poles corresponding to $\Xi(1620)$ in Model~1 and Model~2 originate from different physical mechanisms.

\begin{figure}[tbp]
\centering
\includegraphics[width=7cm]{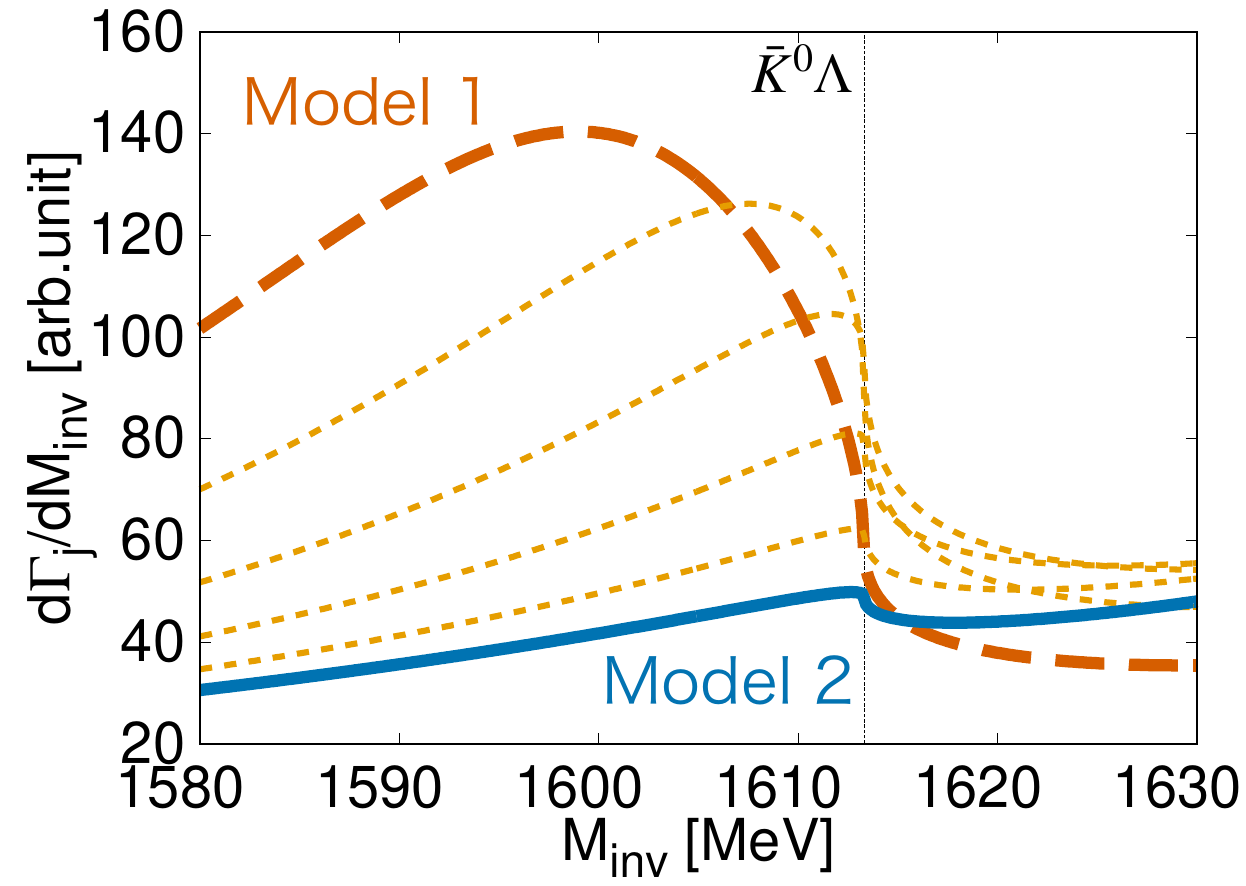}
\includegraphics[width=7cm]{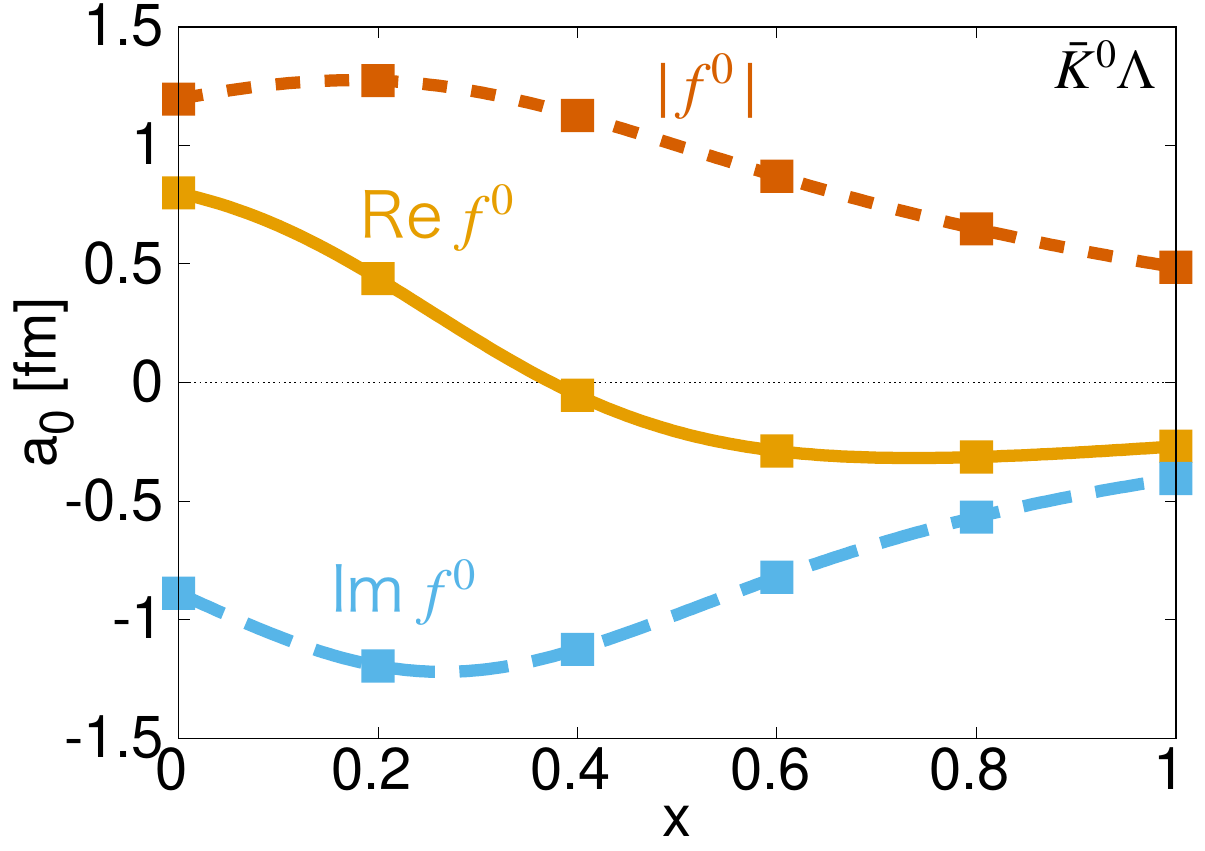}
\caption{Left: Invariant mass distribution of the $\pi^+\Xi^0$ channel obtained via model interpolation. The spectrum for Model~1 ($x=0$) is shown with a dashed line, that for Model~2 ($x=1$) with a solid line, and intermediate spectra for $x$ varied in steps of 0.2 with dotted lines.
Right: Real part (solid line), imaginary part (dashed line), and absolute value (dotted line) of the $\bar{K}^0\Lambda$ scattering length obtained via model interpolation. The scattering lengths corresponding to intermediate values of $x$ (in steps of 0.2) are indicated by squares.
}
\label{fig:spectrum_a0}
\end{figure}

Next, the left panel of Fig.~\ref{fig:spectrum_a0} shows the results of applying the model interpolation to the $\pi^+\Xi^-$ invariant mass distribution in the decay $\Xi_c \to \pi \pi \Xi$, particularly near the $\bar{K}\Lambda$ threshold, which lies relatively close to the $\Xi(1620)$ resonance. The spectra are plotted for values of the interpolation parameter $x$ varied from 0 to 1 in steps of 0.2. As the model parameters are continuously varied from Model~1 to Model~2, the peak in the spectrum gradually approaches the $\bar{K}\Lambda$ threshold. Once the peak reaches the threshold, the spectral maximum progressively decreases and eventually develops into a cusp-like structure.

It is known that the shape of such spectra is strongly influenced by the sign of the real part of the scattering length in the corresponding $\bar{K}\Lambda$ channel~\cite{Dong:2020hxe,Sone:2025xuh}. To clarify this correlation, the right panel of Fig.~\ref{fig:spectrum_a0} shows the variation of the scattering length $a_0$ during the model extrapolation. The real part of the scattering length (solid line) changes sign from positive to negative around $x \sim 0.375$. Examining the corresponding spectra in the left panel, one can clearly observe that the shape of the distribution near the $\bar{K}\Lambda$ threshold changes from a peak to a cusp-like structure around the same value of $x$.

Furthermore, focusing on the spectrum with a clear peak structure at $x = 0$ (Model~1) shown in the left panel of Fig.~\ref{fig:spectrum_a0}, the peak position in the $\pi^+\Xi^-$ distribution in the three-body decay process is found to be $1599~\text{MeV}$. This value is clearly lower than the real part of the pole position in Model~1 ($1610~\text{MeV}$), indicating that the peak is sizably shifted toward lower energy due to threshold effects. A similar threshold effect has been investigated in Ref.~\cite{Nishibuchi:2025uvt}, where the peak position in the two-body scattering amplitude was found to be $1606~\text{MeV}$. This comparison reveals that the threshold effect appears even more prominently in the three-body decay process than in the two-body process.

As shown in Fig.~\ref{fig:pole_trajectory}, all poles plotted at $x \neq 0$ in the interpolation trajectory (evaluated at intervals of 0.2) are located in the energy region above the $\bar{K}\Lambda$ threshold. However, as seen in the left panel of Fig.~\ref{fig:spectrum_a0}, a clear peak structure is still present in the spectrum for $x = 0.2$. This result suggests that strong threshold effects are at play in the three-body decay process. It implies that even when a pole lies above the $\bar{K}\Lambda$ threshold in terms of its complex energy, it can still manifest as a peak below the threshold in the observable spectrum due to such effects.

\section{Summary}

In this study, we have investigated how different eigenstates influence the spectrum associated with the $\Xi(1620)$ resonance in the three-body decay $\Xi_c \to \pi\pi\Xi$, using theoretical models developed in connection with the Belle and ALICE experiments. In Section~2, we have provided an overview of the chiral unitary approach employed in the analysis, the formulation of the model interpolation between the Belle- and ALICE-based models, and the theoretical framework for describing the $\Xi_c \to \pi\pi\Xi$ decay. In Section~3, we focused on the behavior of the poles during the model interpolation and examined in detail how these pole trajectories affect the spectrum near the $\bar{K}\Lambda$ threshold.

As a result, we find that the spectral shape near the $\bar{K}\Lambda$ threshold is strongly correlated with the sign of the real part of the $\bar{K}\Lambda$ scattering length. Furthermore, the threshold effects previously observed in two-body scattering amplitudes are found to appear even more prominently in the invariant mass distribution of the three-body decay process. These findings indicate that, in order to theoretically analyze the $\pi\Xi$ invariant mass spectrum observed in the Belle experiment, it is essential to properly account for the strong threshold effects inherent in the three-body decay amplitude.


\end{document}